\documentclass{ws-p9-75x6-50}
\usepackage{graphicx}
\usepackage{amssymb}

\begin{document}
\title{\normalsize\bf Shot Noise and Proximity Effect in
  Superconductor--Normal-Metal Heterostructures}
\author{W. Belzig, J. B\"orlin and C. Bruder}
\address{Department of Physics and Astronomy, Klingelbergstr. 82,
  University of Basel,\\ 4056 Basel, Switzerland}
\author{Yu. V. Nazarov}
\address{Department of Applied Physics, Delft University of Technology, The
  Netherlands}

\maketitle

\abstracts{ Current noise provides useful information on correlations
  important for transport properties of mesoscopic systems. Of
  particular interest, both experimentally and theoretically, are
  heterostructures of normal metals and superconductors. We use an
  extended Keldysh-Green's function approach to calculate the current
  correlations fully accounting for the superconducting proximity
  effect. The shot noise in diffusive wires shows a reentrant behaviour,
  similar to the reentrance effect of the conductance. At intermediate
  energies the effective charge (the ratio of the differential noise and
  differential conductance) is suppressed below the value of the
  Cooper pair charge 2e due to higher correlations. In superconducting
  heterostructures with more than two normal leads, we address the
  question of the sign of correlations between currents in different
  terminals. We show that positive crosscorrelations are a generic
  feature in these structures, which should be easily observable
  experimentally.}

\section{Introduction}
\label{sec:intro}

The current in mesoscopic structures fluctuates in time. Due to the
coherent nature of the transport, these fluctuations provide information
on the quantum physics of the underlying transport mechanism (see
Ref.~\cite{blanter} for a summary). For example, measuring the current
noise $S_I$ in a tunnel junction allows to extract the effective charge
of the carriers involved. The effective charge of a tunnel junction is
defined similar to Fano factor by the ratio $q_{\rm{eff}}^{\rm
  tun}=S_I/2I$, where $I$ is the average current. This has been used to
determine effective charges of $e/3$ and $e/5$ in the fractional quantum
Hall regime. For non-opaque junctions the definition of an effective
charge only makes sense in reference to something known. For example, in
superconducting heterostructures one usually takes the normal state as
reference point and we may define an effective charge in reference to
that. In this way an effective charge $2e$ in the case of Andreev
reflection for a diffusive contact has been predicted theoretically and
determined experimentally \cite{2e,kozhevnikov,2e-measure}.

It is also possible to measure {\em nonlocal} current-current
correlations, the so-called crosscorrelations.  A fermionic version of
the Hanbury-Brown and Twiss experiment was performed and showed negative
crosscorrelation \cite{hantwissexp}.  These originate from the Pauli
exclusion principle, which leads to a noiseless stream of incoming
particles at zero temperature. The electrons are scattered at a beam
splitter one by one.  Obviously an electron leaving in one lead can not
leave in the other, therefore the time-dependent fluctuations in the two
leads are anticorrelated, thus leading to negative crosscorrelations.
This argument has been put on solid ground by B\"uttiker
\cite{buettiker:91}, who has shown that this holds for transport of
fermions in arbitrary multi-terminal structures with uncorrelated leads.

A natural question arising in this context is, what happens if the incoming
particles are correlated. This can be achieved by using a superconducting
terminal, from which particles are injected by Andreev reflection, i.~e. as
correlated electron-hole pairs. In a single-mode beam-splitter geometry this
can eventually lead to positive crosscorrelations\cite{martin}. Some doubts
have been raised, that the positive correlations survive in the many channel
limit\cite{gramesbacher}. Thus, we will study below one example of a
superconducting beam splitter, in which large positive crosscorrelations
appear\cite{boerlin1}. Similar results have been obtained in different
geometries\cite{taddei2,samuelsson-prl}.


\section{Diffusive Wire}
\label{sec:diff}

\begin{figure}[tbp]
  \begin{center}
    \includegraphics[width=20pc,clip=true]{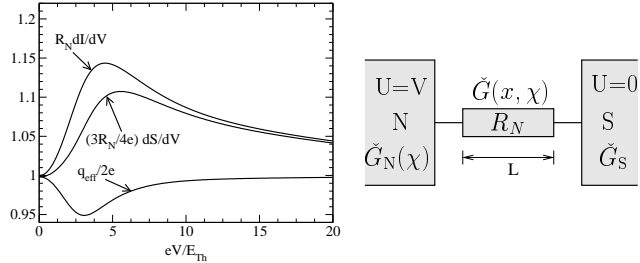}
  \end{center}
  \caption{Transport properties of a diffusive proximity wire. The
    differential conductance $dI/dV$ normalized by the normal state
    resistance shows the usual reentrance behaviour around voltages of
    the order of the Thouless energy $E_{\rm Th}=D/L^2$. The
    differential current noise $dS/dV$ is normalized to $4e/3R_N$, i.e.
    twice the current noise in the normal state. It shows a similar
    reentrance effect as the conductance. The trivial effect of the
    energy dependent conductance on the noise is eliminated in the
    effective charge $q_{\rm eff}= (3/2) dS/dI$. The dip in the
    effective charge below $E_{\rm Th}$ is therefore solely due to
    higher correlations. The right graph shows schematically the proximity
    wire.}
  \label{fig:wire}
\end{figure}

We study a two-terminal geometry, where a diffusive wire is mounted between a
normal and a superconducting terminal, see the inset of Fig.~\ref{fig:wire}.
We expect a doubling of the effective charge, both in the quantum coherent
regime at zero energy, and in the incoherent regime $E\gg E_{Th}$ (with
Thouless energy $E_{Th}=D/L^2$).  However, it is well know that the
conductance is enhanced at intermediate energies due to the proximity effect
-- the so-called reentrant behaviour. To access the noise in this regime, we
make use of an extended Keldysh Green's function method outlined in
Ref.~\cite{belzig1}.  Inside the mesoscopic wire the quasiclassical transport
equations are obeyed\cite{eilenberger}. In a diffusive normal metal wire they
read
\begin{equation}
  \label{eq:usadel}
  D \frac{\partial}{\partial x} 
  \left(\check G(x,\chi)\frac{\partial}{\partial x}\check G(x,\chi)\right) =
  \left[-iE\bar\tau_3\,,\,\check G(x,\chi)\right]\, .
\end{equation}
Here $D$ is the diffusion constant and $x$ the coordinate along the
wire, which has a length $L$.  It's conductance is $G_{\rm{N}} = \sigma
A/L$ (cross section $A$). At both ends boundary conditions to reservoirs
have to be supplied. At the normal end with ideal connection the Green's
function is continuous: $\check G(0,\chi)=\check G_{\rm{N}}(\chi)$. The
current-correlations are encoded in the dependence on the so-called
counting field $\chi$.  The other end is connected to a superconducting
reservoir by a contact of negligible resistance, which leads to the
boundary condition $\check G(L)=\check G_{\rm{S}}$. Details of the
calculation can be found in Ref.\cite{belzig1}.

In Fig.~\ref{fig:wire} we present results for the transport properties at zero
temperature and for $E_{Th}\ll \Delta$. Both the differential current noise
$dS/dV$ and the differential conductance $dI/dV$ show a reentrant behaviour.
The energy dependence of both quantities is nonetheless different,
demonstrating that the noise can not be simply obtained by multiplying a
doubled energy-independent noise $4eG_N V/3$ by the differential resistance.
Consequently this leads to an effective charge, defined by $ q_{\rm{eff}}(V)=
(3/2) dS/dI$, which drops below the value of $2e$ at energies of the below
$\approx 5 E_{Th}$.  We believe this drop is due to a {\em pair-pair}
correlation effect. This is supported by the fact the the suppression of the
effective charge is mainly {\em below} the energy at which the maximal
enhancement of the conductance occurs.  A confirmation of this conjecture
could be obtained in a system with two superconducting terminals, in which
pair-pair correlations could be suppressed by destructive interference.

\section{Beam Splitter}
\label{sec:splitter}

\begin{figure}[t]
  \begin{center}
    \includegraphics[width=20pc,clip=true]{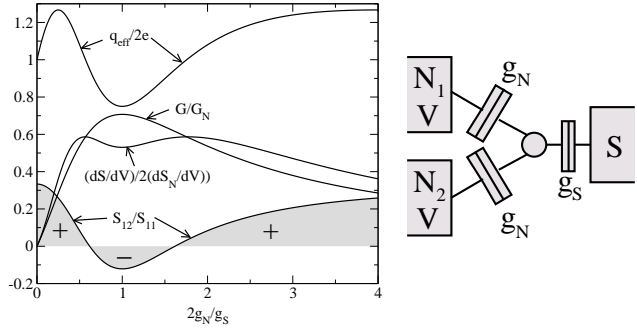}
    \caption{Transport properties of the beam splitter shown on the right. 
      Remarkably, the crosscorrelation for the superconducting
      beam-splitter are positive in most of the parameter range. This is in
      contrast to the normal case, where the crosscorrelations are negative
      for the whole parameter range (not shown). Conductance and noise in the
      superconducting case (normalized to the normal case) show show a
      qualitative similar behaviour like the diffusive wire. However, the
      effective charge, here defined by the ratio $q_{\textrm{eff}}=S I^N /
      S^N I$ shows a non-monotonic behaviour as function of the conductance
      ratio.}
    \label{fig:splitter}
  \end{center}
\end{figure}

To address the question whether positive crosscorrelations can exist in a beam
splitter geometry with one superconducting lead and two normal leads we study
the layout shown in the inset of Fig.~\ref{fig:splitter}.  The superconductor
is held at zero voltage and the two normal terminals are biased symmetrically
at the same voltage $V$. The terminals are connected by tunnel junctions to a
central node. This could be either a small metallic island or a chaotic cavity
and is assumed to be so large, that we can neglect charging effects.  The
transport properties are easily obtained using the circuit theory
formulation\cite{circuit,bagrets}. In fact, for a setup where an arbitrary
number of terminals is connected to one common node by tunnel junctions the
general solution can be obtained\cite{boerlin1}. Note also, that this solution
includes not only the noise, but also the full counting statistics of this
mesoscopic structure\cite{yuli-annals}.

We will now concentrate on auto- and crosscorrelations only. These are
defined as $S_{ij}=2\int dt \langle \delta I_i(t) \delta I_j(0)\rangle$,
where $I_{i(j)}(t)$ are the time dependent currents in terminal $i(j)$. We
obtain in the low-energy limit ($T=0$, $V\ll \Delta$):
\begin{eqnarray}\nonumber
  G = \frac{g_S^2g_N^2}{(g_S^2+g_N^2)^{3/2}}\qquad&,&
  S = 2G|eV| 
  \left(2-5\frac{g_S^2g_N^2}{(g_S^2+g_N^2)^{2}}\right)\,,
  \\\nonumber
  S_{12} = \frac{G|eV|}{2}
  \left(1-5\frac{g_S^2g_N^2}{(g_S^2+g_N^2)^{2}}\right)&,&
  S_{11} = \frac{G|eV|}{2}
  \left(3-5\frac{g_S^2g_N^2}{(g_S^2+g_N^2)^{2}}\right)\,.
\end{eqnarray}
For completeness we also cite the results for the case, in which the
superconductor is in its normal state:
\begin{eqnarray}\nonumber
  G^N =  \frac{g_Sg_N}{(g_S+g_N)}\qquad&,&\quad
  S^N =  2G^N|eV| \left(1-2\frac{g_Sg_N}{(g_S+g_N)^2}\right)
  \,,\\\nonumber
  S_{12}^N = -\frac{G^N|eV|}{2} \frac{g_Sg_N}{(g_S+g_N)^2}&,&\quad
  S_{11}^N = \frac{G^N|eV|}{2} \left(2-3 
  \frac{g_Sg_N}{(g_S+g_N)^2}\right)\,.
\end{eqnarray}
In Fig.~\ref{fig:splitter} we summarize the transport properties,
emphasizing the difference between the superconducting state and the
normal state. Note that the total current noise is given by
$S=2S_{11}+2S_{12}$. An interesting observation is that the transport
properties are invariant under inversion of the conductance ratio
$g_N/g_S$. In the following we will thus distinguish the case of weak
(or strong) proximity effect determined by $g_N\not\approx g_S$ and the
case of optimal proximity effect defined by $g_N\approx g_S$. Below we
will use the terminology weak for both the weak and the strong proximity
regimes equally.

The conductance vanishes rapidly in the limit of weak proximity and
shows a resonance in the regime of optimal proximity. This actually
defines the condition for optimal proximity effect: the regime, in which
the proximity effect has the largest impact on the conductance. Note,
that without proximity effect the conductance would vanish always in the
present geometry. In Figure \ref{fig:splitter} we defined the effective
charge in the superconducting case with respect to the normal case:
$q_{\textrm{eff}}=S I^N / S^N I$. This is in analogy to the diffusive
wire, studied in the previous section. Remarkably, the effective charge
shows an \textit{increase} above $2e$ for the case of weak proximity.
Around the resonance condition $g_S\approx g_N$ the effective charge
drops below $2e$ and reaches a minimum of $(3/2)e$ for $g_S=g_N$. We note
that this behaviour stems from our definition of the effective charge.
The Fano factor for the superconducting case, defined by $F=S/2eI$,
never increases above 2.
 
The most surprising observation is that we can obtain positive
crosscorrelations in the superconducting beam splitter, whereas they are
manifestly negative for the normal beam splitter. Interestingly the positive
crosscorrelations are found in the limit of weak proximity\cite{boerlin1,samuelsson-prl}, or in the absence of
proximity\cite{samuelsson-prep}.  Thus, they should be obtained from
simple arguments. Indeed, from a calculation of the full counting
statistics\cite{boerlin1}, it follows that in the weak proximity limit the
statistics consists of \textit{independent} events of pair tunneling. The
possible events are pair tunneling into terminal $N_1$ or $N_2$ and correlated
tunneling in both terminals.  These events occur with equal probabilities,
however, in the limit of weak proximity these events are
\textit{uncorrelated}.  Thus, two particle tunneling events into the same lead
do \emph{not} contribute to cross correlations. In contrast, tunneling of two
particles into different leads is \emph{automatically} positively correlated.
Thus, it follows quite generally, that the crosscorrelations are positive.

Following this argumentation to the regime of optimal proximity effect
in fact shows, that it is much harder to understand the occurrence of
\emph{negative} correlations in this regime.  We propose the following
interpretation. These negative crosscorrelations occur in the same
regime, in which the effective charge drops below $2e$. For uncorrelated
electrons a suppression of the effective charge is interpreted as a
consequence of the Fermi statistics. More precisely, it is a consequence
of the Pauli principle, that two fermions can not occupy the same state.
Now, the same holds obviously for pairs of electrons. If a two particle
state is occupied by two electrons, this state is blocked for other
pairs of electrons. Cooper pairs are such objects, and can not occupy the
same state twice. Note, that we speak of Cooper pairs, but rather mean
an correlated electron pair, which has entered the normal metal node. In
the limit of weak proximity, the average occupation of a state on the
central node with an electron pair is small, and the occupation does not
matter, therefore. The same holds in the limit of strong proximity, where
the state is nearly always occupied. This changes in the limit of
optimal proximity, where the occupation approaches $1/2$. Just as in the
case of Fermions, this leads to a strong suppression of the Fano
factor.

To understand the negative crosscorrelations we follow the previous
argumentation. The Pauli repulsion of Cooper pairs reduces the noise in
the 'incoming' beam of Cooper pairs (similar to lowering the temperature
in a fermionic beam). Let us, therefore, consider an incoming
'noiseless' beam of Cooper pairs, which is distributed equally among the
possible outgoing states. The crosscorrelations can be written in terms
of occupations of the outgoing states $n_{1,2}$ as $s_{12}=\langle
n_1n_2 \rangle-\langle n_1 \rangle\langle n_2 \rangle$. The possible
outgoing states are $(n_1=2,n_2=0)$, $(n_1=0,n_2=2)$, and
$(n_1=1,n_2=1)$, where the last state is doubly degenerate. The average
occupation in terminal 1 is then $\langle n_1 \rangle= (1/4)\times 2 +
(1/2)\times 1= 1$, and the same for terminal 2. For the
crosscorrelations we need $\langle n_1n_2 \rangle$. The first two
outgoing states do not contribute, whereas the third yields $(1/2)\times
1$. Collecting all terms we find $s_{12}=-1/2$, which is negative. The
value of the crosscorrelations in Fig.~\ref{fig:splitter} is only
$-1/8$. This discrepancy is probably due to the finite backscattering in
the double tunnel junction geometry. Nevertheless the qualitative
behaviour is correct. 

\section{Summary}
\label{sec:summ}

We have shown that current-correlations in multi-terminal
heterostructures display a number of quite remarkable properties, which
depend on the phase coherent nature of quantum transport. The proximity
effect in a diffusive wire modifies the conductance and the noise in a
nontrivial way. Both are enhanced at energies of the order of the
Thouless energy, which divides coherent and incoherent Andreev
transport. We have defined an effective charge by referencing to the
noise in the normal state. The effective charge displays a suppression
below $2e$ for energies below $E_{Th}$, which could be due to coherent
multiple Andreev-pair transport.
For a multi-terminal structure with one superconducting and two normal
metal arms, we have shown that positive crosscorrelations of the
currents in the two arms are a generic feature in these system. It is
remarkable, that in the region, in which the proximity effect has the
strongest impact on the conductance, the crosscorrelations are
negative. This can be explained by a Pauli exclusion principle for
Cooper pairs in a transport process.
Finally, we would like to emphasize the similarity of the noise
behaviour in a diffusive wire and a multiple tunnel junction geometry.
In particular, the suppression of the Fano factor in the regime of
optimal proximity effect has probably the same origin. For the
tunnel-junction geometry the finite occupation of the central node with
Cooper pairs leads to a reduces Fano factor and positive cross
correlations. In the diffusive wire the effective charge (equivalent to
the Fano factor) are suppressed below energies of the order of $E_{Th}$,
which is probably due to a finite occupation of the wire with
Cooper pairs. A deeper understanding of this phenomenon is obviously
interesting for further studies.

We acknowledge useful discussions with M. B\"uttiker, J.~C. Cuevas, A.~A.
Kozhevnikov, D.~E. Prober, B. Reulet, and P. Samuelsson. This work was supported by the Swiss
NSF and the NCCR "Nanoscience".


\begin{thebibliography}{}
\bibitem{blanter}  
  Ya.~M. Blanter and M. B\"uttiker, Phys. Rep. {\bf 336}, 1 (2000).
\bibitem{2e} V.~A. Khlus, Sov. Phys. JETP {\bf 66}, 1243 (1987); B.~A.
  Muzykantskii and D.~E. Khmelnitskii, Phys. Rev. B {\bf 50}, 3982 (1994);
  M.~J.~M. de Jong and C.~W.~J. Beenakker, \textit{ibid.} {\bf 49}, 16070
  (1994); K.~E.  Nagaev and M. B\"uttiker, \textit{ibid.} {\bf 63}, 081301(R)
  (2001).

\bibitem{kozhevnikov}
 A.~A. Kozhevnikov \textit{et al.}, 
 Phys. Rev. Lett {\bf 84}, 3398 (2000).

\bibitem{2e-measure}
  X. Jehl \textit{et al.}, Nature {\bf 405}, 50 (2000).

\bibitem{hantwissexp} M. Henny {\textit et al.}, Science {\bf 284},
 296 (1999); W.~D. Oliver \textit{et al.}, Science {\bf 284}, 299
 (1999); S. Oberholzer \textit{et al.}, Physica (Amsterdam) E {\bf 6},
 314 (2000).

\bibitem{buettiker:91}
 M. B\"uttiker, Phys. Rev. Lett. {\bf 65}, 2901 (1990).
 
\bibitem{martin} T. Martin, Phys. Lett. A {\bf 220}, 137 (1996); M.~P.
 Anantram and S. Datta, Phys. Rev. B {\bf 53}, 16390 (1996).

\bibitem{gramesbacher}
 T. Gramespacher and M. B\"uttiker, Phys. Rev. B {\bf 61}, 8125 (2000).

\bibitem{boerlin1}
  J. B\"orlin, W. Belzig, and C. Bruder, Phys. Rev. Lett. {\bf 88}, 197001 (2002).

\bibitem{taddei2} F. Taddei and R. Fazio, Phys. Rev. B {\bf 65}, 134522 (2002).
  
\bibitem{samuelsson-prl}
  P. Samuelsson, M. B\"uttiker, Phys. Rev. Lett. {\bf 88},  046601 (2002).

\bibitem{belzig1}
  W. Belzig and Yu. V. Nazarov, Phys. Rev. Lett. {\bf 87}, 067006 (2001). 

\bibitem{eilenberger}
  A.~I. Larkin and Yu.~N. Ovchinnikov,
  Sov. Phys. JETP {\bf 26}, 1200 (1968);  
  K.~D. Usadel, 
  Phys. Rev. Lett. {\bf 25}, 507 (1970).
  
\bibitem{circuit}Yu.~V. Nazarov, Superlattices Microst. {\bf 25}, 1221 (1999).

\bibitem{bagrets}
 Yu.~V. Nazarov and D. Bagrets, Phys. Rev. Lett. {\bf 88}, 196801 (2002).

\bibitem{yuli-annals}
 Yu.~V. Nazarov, 
 Ann. Phys. (Leipzig) {\bf 8}, SI-193 (1999).
 
\bibitem{samuelsson-prep} P. Samuelsson and M. Buttiker, cond-mat/0207585
  (unpublished).


\end{thebibliography}
\end{document}